\documentclass[reprint,amsmath,amssymb,aps,prl,showpacs]{revtex4-1}

\usepackage{graphicx}% Include figure files
\usepackage{dcolumn}% Align table columns on decimal point
\usepackage{bm}% bold math
\let\ltxRightarrow\Rightarrow
\usepackage{marvosym}
\let\Rightarrow\ltxRightarrow
\usepackage{wasysym}
\usepackage{amssymb}
\usepackage{color}
\definecolor{gray}{rgb}{0.5, 0.5, 0.5}

\begin{document}

\preprint{APS/123-QED}

\title{Frequency Dependence of the Supersolid Signature in Polycrystalline $^4$He}% Force line breaks with \\

\author{George Nichols}
\author{Malcolm Poole}
\author{Jan Ny\`eki}
\author{John Saunders}
\author{Brian Cowan}
\email{b.cowan@rhul.ac.uk}
\affiliation{Department of Physics, Royal Holloway University of London, Egham, Surrey, TW20 0EX, UK.}%Lines break automatically or can be forced with \\

\date{September 19, 2014}% It is always \today, today,

\begin{abstract}

 We report studies, using a two mode torsional oscillator, of the putative supersolid signature in polycrystalline
$^4$He. Measurements at two frequencies enable us to eliminate the viscoelastic contribution to the signature, and other instrumental effects arising from the temperature dependent shear modulus of the sample. The complex response function of the sample, encoded \emph{via} its effective moment of inertia, shows an unexpected and unexplained frequency dependence. This cannot be accounted for by glassy dynamics within the sample. The results do not rule out the possibility of supersolidity in bulk solid $^4$He.

\end{abstract}

\maketitle

%\tableofcontents

The supersolid state of matter corresponds to the coexistence of broken gauge symmetry and broken translational symmetry. This quantum state challenges our fundamental understanding of the essence of solidity. It was originally proposed theoretically that it could be realised in bulk solid $^4$He \cite{Andreev1969, Chester1970, Leggett1970}, and the observations by Kim and Chan \cite{Kim2004} triggered an explosion of both experimental and theoretical activity, and significant controversy. We refer to \cite{Chan2008, Prokofev2007, Balibar2010, Boninsegni2012, Galli2008} for recent reviews.    

In Leggett's original discussion of supersolidity \cite{Leggett1970} he suggested it might be observed by detection of a non-classical rotational inertia (NCRI) in a DC rotation experiment, along the lines of the Hess-Fairbank \cite{Hess1967} experiment on superfluid $^4$He. The experiments of Kim and Chan \cite{Kim2004}, followed by those of Rittner and Reppy \cite{Rittner2006} and others \cite{Kondo2007, Penzev2007, Aoki2007, Clark2007} were performed in a torsional oscillator (TO), where the specimen is subjected to an angular  \emph{oscillation}. They observed a drop in the oscillator's period which was interpreted as a reduction of the specimen's moment of inertia. This missing moment of inertia (MMI) was identified with NCRI and interpreted to be the signature of a supersolid transition.  

The detection of the putative supersolid response using the torsional oscillator method is challenging, since measurements are performed on a solid, the elastic response of which may mask any small mass decoupling arising from supersolidity. Following the original ``discovery" of supersolidity, the elastic modulus of solid $^4$He, the temperature dependence of which arises from the motion and pinning of dislocations, has been extensively investigated~\cite{Day2007,Day2010,Syshchenko2010,Rojas2010,Haziot2013,Balibar2012,Beamish2012b,Haziot2013b}. A variety of ways sample elasticity can influence observations using torsional oscillators, has been discussed~\cite{Balibar2008,Maris2010,Balibar2012}. Essentially the elasticity can influence the observed oscillator frequency either (i) through its influence on the torsion constant, obscuring the changes we wish to detect in the moment of inertia~\cite{Beamish2012,Maris2012}, or (ii) through the viscoelastic response of the solid helium specimen \cite{Yoo2009,Reppy2012}. However, the magnitude of these effects depends on, and can thus be controlled by, TO design parameters and sample geometry. They also depend on operating frequency. This has motivated the most recent work, using the torsional oscillator technique, to set limits on any possible supersolid signature \cite{Mi2014,Kim2014}.

Other earlier work emphasized the potential importance of the glassy dynamics of crystal defects \cite{Nussinov2007,Hunt2009,Gadagkar2012} in determining sample properties in a TO experiment. This perspective highlighted the critical importance of measuring and interpreting the full complex response function of the sample to oscillatory rotation. This provides a characteristic signature as the glassy component of the sample moves on cooling  between the regimes $\omega\tau\ll 1$ to $\omega\tau\gg1$, where $\tau$ is the relaxation time of the glassy component. Indeed in previous work also, we found that characterizing TO measurements by a parametric plot of the real \textit{vs} imaginary parts of the response (Cole-Cole plot), is a powerful method to fingerprint its underlying mechanisms. The two examples studied were (i) the vortex dynamics at a Kosterlitz-Thouless transition in two dimensional liquid $^4$He \cite{Bowley1998}, first discovered by Bishop and Reppy~\cite{Bishop1978} and (ii) the non-superfluid apparent mass decoupling of a sub-monolayer helium film on graphite~\cite{Mohandas1995}.

In this Letter we report new measurements using a compound torsional oscillator with two modes, widely separated in frequency by a factor of 5, to better disentangle the above competing effects. The sample geometry, a cylinder of diameter 14mm and height 2mm and hence relatively high aspect ratio, is tailored to have a negligible viscoelastic response. Our central result is that we find a frequency dependence of the complex response to angular oscillation of solid $^4$He, characterized by Cole-Cole plots, which is unusual and not explained by elastic response. These results, in the context of studies using a torsional oscillator \emph{and} DC rotation which show an unexplained influence of the rotation on the period drop~\cite{Choi2010, Choi2012, Yagi2011}, add to the list of intriguing observations which may support the existence of a supersolid phase.  

A conventional torsional oscillator operates at a single frequency. It is convenient to express its behaviour in terms of the rotational susceptibility $\chi(\omega)$
\begin{equation}
\chi^{-1}(\omega)=k- \mathrm i \omega\gamma-\omega^2I_\mathrm{osc}-\omega^2I_\mathrm{eff}(\omega)
\end{equation}  
where $I_\mathrm{osc}$ is the oscillator's moment of inertia,  $k$ is the  torsion constant and  $\gamma$ its dissipation coefficient ($\gamma=\omega I_\mathrm{osc}/Q_0 $). Here $Q_0$ is the quality-factor of the empty oscillator of resonant (angular) frequency $\omega_0=\sqrt{k/I_\mathrm{osc}}$. The effect of the helium specimen in the oscillator is embodied in the complex `effective moment of inertia' function $I_\mathrm{eff}(\omega)$.  The real part gives rise to period shifts and the imaginary part to dissipation. Different physical phenomena in the helium specimen will manifest as different functional forms for $I_\mathrm{eff}(\omega)$.    (We note the connection between $I_\mathrm{eff}(\omega)$ and the back action function $g(\omega)$ of~\cite{Nussinov2007}: $g(\omega)=\omega^2 I_\mathrm{eff}(\omega)$).
 
$I_\mathrm{eff}(\omega)$ is determined by measuring  the shift of the period and dissipation from those of a reference state.  In supersolid experiments it is  convenient to choose this reference as full mass loading of the helium specimen. Then the period and dissipation shifts are related to $I_\mathrm{eff}$ by
\begin{equation}\label{gw-expr}
\frac{I_\mathrm{eff}(\omega)-I_\mathrm{He}}{I_\mathrm{osc}}=\frac{\Delta I_\mathrm{eff}(\omega)}{I_\mathrm{osc}}=2\frac{\Delta P}{P}+\mathrm i \Delta\frac{1}{Q}
\end{equation} 
where $I_\mathrm{He}$ is the helium rigid body moment of inertia.

\begin{figure}[h]
\includegraphics[scale=0.17]{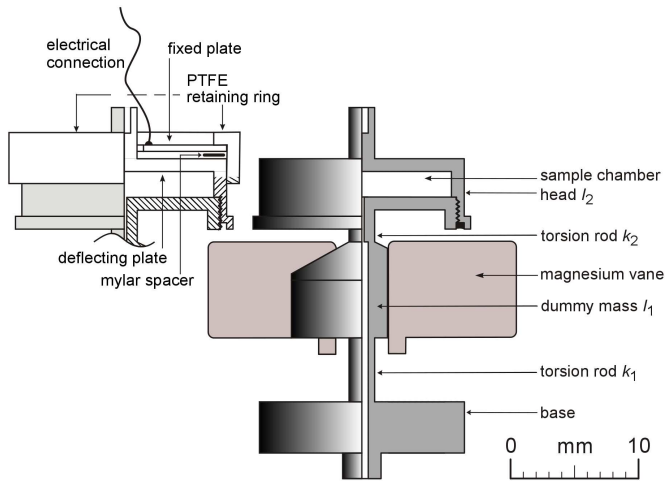}% Here is how to import EPS art
\caption{\label{cellbodyassembled2} Schematic of the double oscillator used for these measurements.}
\end{figure}

Ideally one would like to vary the frequency and thereby trace out the functional form of $I_\mathrm{eff}(\omega)$, in order to identify a potential supersolid response. But a conventional oscillator operates at a single frequency. It is possible, however, to construct a two-mode oscillator, as in Fig.~\ref{cellbodyassembled2}. This method was first applied to the study of solid $^4$He in \cite{Aoki2007}.
Such an oscillator has symmetric and anti-symmetric
oscillation modes with (angular) frequencies $\omega_\mathrm s$ and $\omega_\mathrm a$ given by 
\begin{equation}\label{mode_freqs}
\omega_\mathrm{a/s}^2=\tfrac{1}{2}\Bigg[ \left(\tfrac{k_1}{I_1}+\tfrac{k_2}{I_1}+\tfrac{k_2}{I_2} \right)
\pm\sqrt{\left(\tfrac{k_1}{I_1}+\tfrac{k_2}{I_1}+\tfrac{k_2}{I_2}\right)^2-4\tfrac{k_1}{I_1}\tfrac{k_2}{I_2}}
\Bigg]
\end{equation}
where the oscillator elements have moments of inertia $I_1$ and $I_2$ and the rods have torsion
constants $k_1$ and $k_2$.

In a double oscillator the sensitivity of the period and dissipation shifts to $I_\mathrm{eff}(\omega)$ will be different for the two modes; the analogue of Eq.~\eqref{gw-expr} may be written
\begin{equation}\label{AA}
\begin{split}
\frac{I_\mathrm{eff}(\omega_\mathrm{a/s})-I_\mathrm{He}}{I_2}&=
\frac{\Delta I_\mathrm{eff}(\omega_\mathrm{a/s})}{I_2}\\
&=\frac{1}{2\beta_\mathrm{a/s}}\left(2\frac{\Delta
P_\mathrm{a/s}}{P_\mathrm{a/s}}+\mathrm i \Delta\frac{1}{Q_\mathrm{a/s}}\right)\\
\end{split}
\end{equation}  
where $\beta_\mathrm a = \partial\ln P_\mathrm a/\partial\ln I_2$ and $\beta_\mathrm s = \partial\ln P_\mathrm s/\partial\ln I_2$ are the sensitivity factors,
and
\begin{equation}
\beta_\mathrm a=\frac{(k_2/I_2-\omega_\mathrm s^2)}{2(\omega_\mathrm a^2-\omega_\mathrm s^2)},
\quad \beta_\mathrm s=\frac{1}{2}-\beta_\mathrm a
\end{equation}
follow from  differentiation of Eq.~\eqref{mode_freqs}. In this way we see that by dividing the fractional period shift and dissipation shift measured at the antisymmetric/symmetric mode   by $2\beta_\mathrm{a/s}$ we obtain the period shift and dissipation shift one \emph{would obtain} from a single mode oscillator operating at  frequency $\omega_\mathrm{a/s}$.

Our compound oscillator,  constructed from coin silver, and incorporating an \emph{in situ} pressure  gauge, is shown in Fig.~\ref{cellbodyassembled2}. The pressure was observed through the deflection of the upper surface of the sample chamber, sensed capacitively using a cryogenic back diode oscillator \cite{Degrift1975}. The gauge assembly was held tightly by a PTFE retaining ring, and the mylar spacer secured with vacuum grease. The helium specimen is contained in element 2 and the capacitive drive and pick-up occur through the vanes attached to element 1. The mode frequencies are $\omega_\mathrm a /2\pi =1978.5$ Hz and  $\omega_\mathrm s /2\pi= 399.14$ Hz, a ratio of almost 5.
 At the higher AC drive levels a frequency stability of  about 3 in $10^9$ was achieved at the high frequency mode and about 1 in $10^8$ at the low frequency mode.

The parameters for the double oscillator are experimentally determined, from Eq.~\eqref{mode_freqs}, by filling the cell with solid helium, and measuring the change of frequency of both modes with mass loading. We determine: $I_1 = 1.43  \times 10^{-7}$\,kg\,m$^2$, $I_2 = 3.30  \times 10^{-7}$\,kg\,m$^2$, $k_1=  3.35$\,N\,m, $k_2 =  13.7$\,N\,m. The results from filling with $^4$He and $^3$He are in good agreement.

\begin{figure}[h]
\includegraphics[scale=0.5]{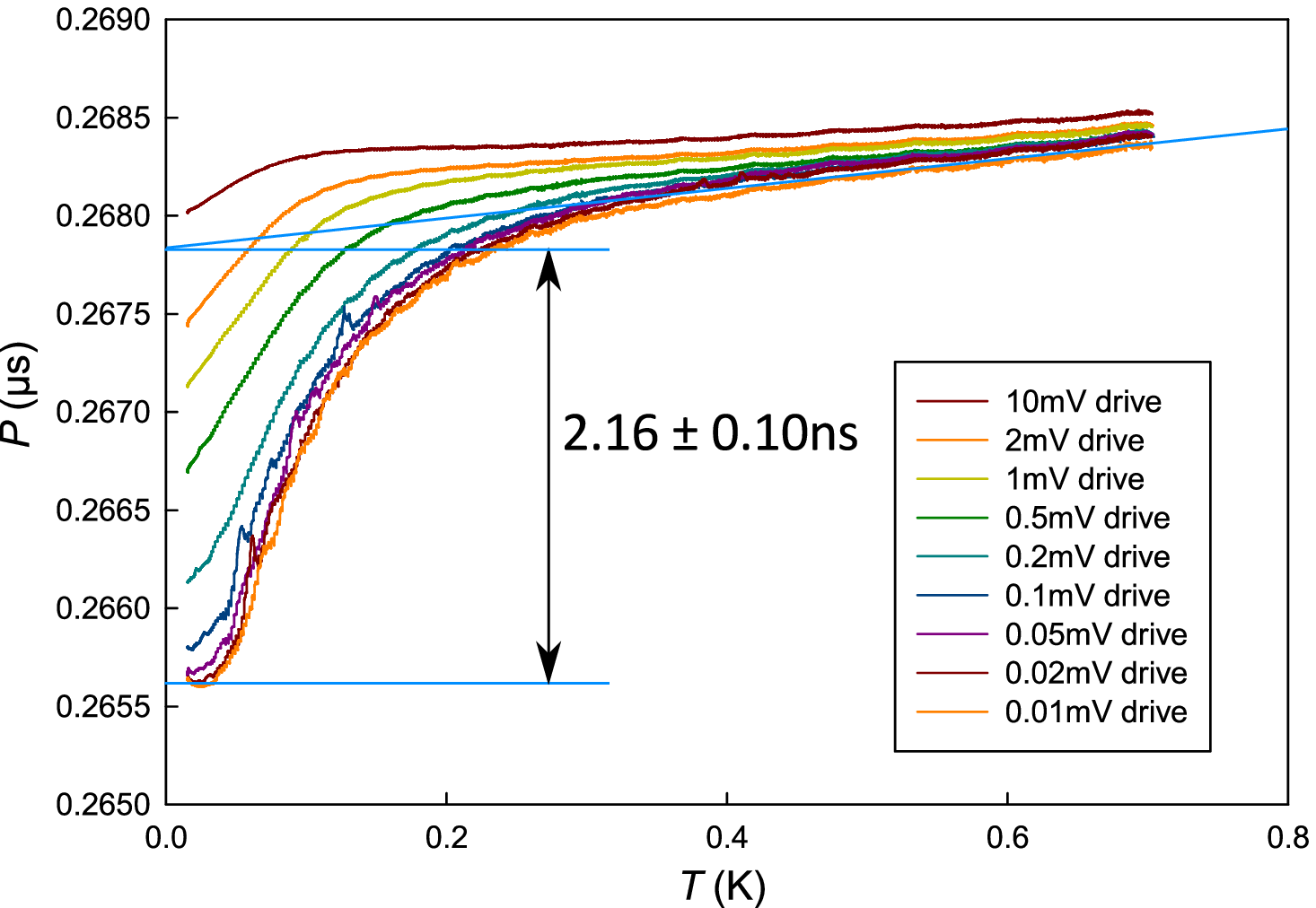}
\includegraphics[scale=0.625]{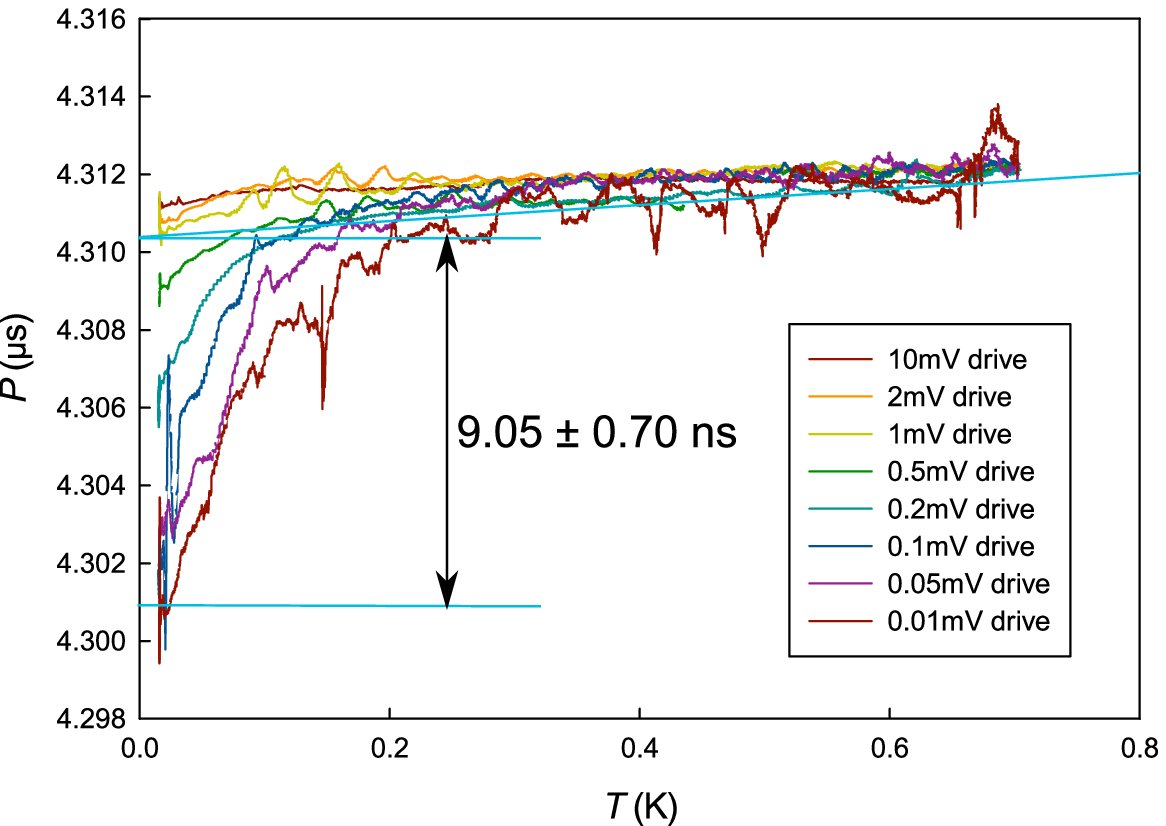}
\caption{\label{hi-low-period} Period data for the high and low frequency modes, with empty cell background subtracted.}
\end{figure}

Measurements were made on a poly-crystalline sample of $^4$He at a pressure of 42~bar, grown by the blocked capillary method. Fig.~\ref{hi-low-period} shows measurements of the high and low mode oscillator period over the temperature range 700~mK to 15~mK for a range of excitation levels, following subtraction of the empty cell background.
We note that for low excitations the period behaviour is independent of drive level \footnote{A comparative analysis of the reduced response at higher drive levels for both modes implies a critical $\it{velocity}$ of 100 $\mu$ms$^{-1}$, as found by \cite{Aoki2007}. See also Supplementary Information. We note that this contrasts with the critical amplitude (strain) effects observed in studies of the nonlinear elastic response \cite{Day2010}.}. There is a gentle period decrease on cooling from the high temperature end and a more rapid drop starting in the vicinity of 200~mK. We determined the period drop of the low-temperature feature by extending a straight line through the higher-temperature data, as shown on the figures, to give the full mass loading periods. In this way we determined the low-drive period drops associated with the low-temperature feature to be $\Delta P_\mathrm{a}=2.16\pm0.10$\,ns and $\Delta P_\mathrm{s}=9.05\pm0.70$\,ns for the antisymmetric and the symmetric modes. 

We can express the apparent change in moment of inertia of the sample as a fraction of the rigid body helium moment of inertia, by multiplying the results of Eq.~\eqref{AA} by $I_2/I_\mathrm{He}=220$, where $I_\mathrm{He}= 1.50\times10^{-9}$\,kg\,m$^2$ is the moment of inertia of our helium specimen, at 42~bar. We obtain  $\Delta I_\mathrm{eff}(\omega_\mathrm a)/I_\mathrm{He}=(8.04\pm 0.37)\times10^{-3}$ and $\Delta I_\mathrm{eff}(\omega_\mathrm s)/I_\mathrm{He}=(2.10\pm 0.16)\times10^{-3}$. 

\begin{figure}[h]
\includegraphics[scale=0.5]{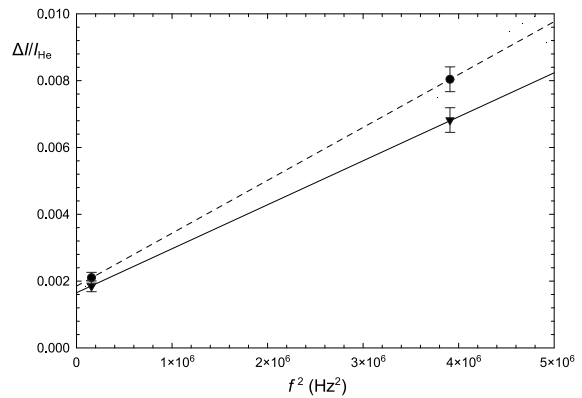}
\caption{\label{freq-squared-plot} Frequency dependence of the missing moment of inertia. Circles show `raw' data and triangles show the result after corrections discussed in the text. }
\end{figure}

Following Reppy \cite{Reppy2012}, we plot the MMI as a function of frequency squared, Fig.~\ref{freq-squared-plot},  since elastic effects have this frequency dependence.
The zero-frequency intercept of the line in such a plot, in this case $(1.86\pm 0.18)\times 10^{-3}$, has been proposed as the signature for the supersolid MMI fraction \cite{Mi2012}. (We should note
that the low frequency point is sufficiently close to the axis to give a good estimate of the zero frequency intercept directly). The double-frequency measurements on solid helium confined in a vycor matrix reported in \cite{Mi2012}, when analysed in this way, show a zero or vanishingly small supersolid MMI, supporting recent measurements by Chan \cite{Kim2012} on such a system and the contention that there is no supersolidity of $^4$He in a vycor matrix. On the other hand a recent measurement \cite{Mi2014} on bulk helium using a double oscillator yields a finite zero-frequency intercept, as found in our experiment.
\begin{figure}[h]
\includegraphics[scale=0.5]{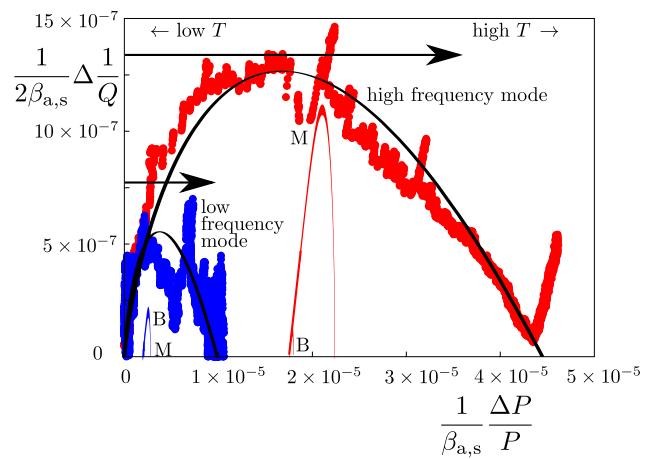}
\caption{\label{fig:Oscillator-Data} Cole-Cole plots of the complex $I_\mathrm{eff}(\omega)$ from the dissipation and period shift data. The solid curves correspond to the special Havriliak-Negami
function, discussed in the text below. The arrows show the inferred MMI period shifts for the two modes. The red and blue curves marked M and B, are the estimated corrections due to changes in the effective torsion constant, discussed in \cite{Beamish2012,Maris2012} (see Supplementary Information), based on the data of \cite{Syshchenko2010}. }
\end{figure}

One potential source of the squared frequency dependence is the viscoelastic response of the solid helium specimen. This leads to a frequency-dependent effective moment of inertia \cite{Yoo2009,Reppy2012}
\begin{equation}\label{VE}
I_\mathrm{eff}(\omega)=I_0\left(1+\omega^2\frac{\rho R^2 }{24\mu}\times F(R/h)  \right),
\end{equation}
where $R,h$ are the radius and height of the cylindrical sample.
$\rho$ is the helium density and $\mu$ its
shear modulus. $F(R/h)$ takes the value 0.112 for our high aspect ratio sample geometry.  A stiffening of the helium in the cell, will affect $I_\mathrm{eff}(\omega)$ directly, as in Eq.~\eqref{VE}. This will lead to oscillator period shifts and dissipation, Eq.~\eqref{AA},
since $\mu$ is complex. Beamish \textit{et al.} \cite{Day2007} observed that the shear modulus of solid $^4$He increases significantly below 200\,mK. The 8\% measured values for the stiffening found in \cite{Syshchenko2010}  would cause  an increase in $\Delta P_\mathrm s$ of 6.72\,ps and
an increase in $\Delta P_\mathrm a$ of 10.3\,ps, which are negligible. The corrections to the MMI  arising from the temperature dependence of the shear modulus on the effective torsion constant \cite{Beamish2012,Maris2012} are discussed in the Supplementary Information; they are shown in Fig.~3. These corrections are relatively modest (the scale of these effects could be reduced still further by improved oscillator design). The finite zero frequency intercept of the MMI plotted as a function of squared frequency is thus a robust result. It is consistent with a supersolid response, but could, in principle, have other explanations e.g. glassy dynamics. 

Therefore, as previously discussed, it is essential to evaluate the complex response function  $I_\mathrm{eff}(\omega)$, and the potential contributions to it. The data at both frequencies are presented in Fig.~\ref{fig:Oscillator-Data}, as a Cole-Cole plot. The estimated contributions of the effects discussed in \cite{Beamish2012,Maris2012}, modelled using results for the complex shear modulus \cite{Syshchenko2010} (see Supplementary Information) are shown at both frequencies and are modest. 

The key result we must account for is the striking difference of the Cole-Cole plots for the two frequencies \footnote{There are possibly hints of this in the analysis of the double oscillator experiment of \cite{Aoki2007}, by \cite{Graf2011}. However the TO data is subject to large corrections from elastic effects, see Fig. 1 in \cite{Beamish2012}.}.  We find that a self-consistent explanation of the frequency dependence, under the hypothesis that the response arises from glassy dynamics, is not possible. To model the glassy dynamics of   the $^4$He sample, we can introduce an effective moment of inertia, which might take the Cole-Cole \cite{Cole1941} form $I(\omega ) = {I_0}/[1 + {(i\omega \tau )^\alpha}]$ or the Davidson-Cole \cite{Davidson1951}  form $I(\omega ) = {I_0}/{[1 + i\omega \tau ]^\gamma}$. We also consider a generalization of these, the the special Havriliak-Negami
 \footnote{The Havriliak-Negami function \cite{Havriliak1966} is a two-parameter generalization of the Cole-Cole and the Cole-Davidson functions with the form $[1+(i\omega\tau)^\alpha]^{-\gamma}$. The \emph{special} form corresponds to the limiting case $\gamma=1/\alpha$ \cite{Havriliak1994}.} form $I(\omega)=I_0/[1+(i\omega\tau)^s]^{1/s}$. The features of our plots rule out the first two: the asymmetry of the plots is incompatible with the Cole-Cole form which describes the chord of a circle,   while the Davidson-Cole form cannot accommodate the asymmetry \emph{and} the aspect ratio of the data. We adopt the special Havriliak-Negami form. This (as do all these forms) predicts that the plots for the two modes collapse onto a single curve, which is clearly not the case. A choice of exponent $s=1$, corresponding to simple Debye relaxation with a single relaxation time, predicts an aspect ratio of the imaginary to real part of the Cole-Cole plot of 0.5, far away from the observed values of 0.04 for the low frequency mode and 0.03 for the high frequency mode. In other words the frequency shift is much larger than the corresponding dissipation maximum of this simple model, as found previously at a single frequency \cite{Hunt2009}; the Cole-Cole plot is highly stretched along the normalised period shift axis.  While this shape can be fitted at each frequency by adjusting the exponent $s$, no consistent description can be found for the two modes; we find exponents $s=0.05$ and 0.07 for the high and low modes respectively -- these are the solid lines through the data in Fig.~\ref{fig:Oscillator-Data}.

 The exponent $s$ may be regarded as characterizing a distribution of relaxation times \cite{Bertelsen1974}. Our values, close to zero, are striking; they correspond to a very flat distribution in ln $\tau$. In stark contrast, we find that if the complex shear modulus data \cite{Syshchenko2010} are plotted on a Cole-Cole plot (see Supplementary Information), they are well described by the Davidson-Cole form with an exponent $\gamma=0.22$. An analysis \cite{Su2010} of the same data has shown that the Cole-Cole plots of measurements at 2, 20 and 200 Hz collapse, a property of both Cole-Cole or Davidson-Cole distributions. 
The non-collapse of the torsional oscillator response at two frequencies, and the different functional form for the Cole-Cole plot, strongly argues that a different mechanism governs the TO response \footnote{We can introduce an  \textit{ad hoc} frequency dependence by multiplying the effective moment of inertia by a factor ${\left( {\omega /\omega_0 } \right)^q}$. For our measurements an index $q=0.82$ would provide the appropriate \emph{horizontal} scaling of the plots of Fig.~\ref{fig:Oscillator-Data}, but not map the Cole-Cole plots onto each other.}.
  
In the light of these results and analysis, we make the \emph{conjecture} that the strong frequency response observed is an intrinsic property of supersolidity in bulk solid $^4$He at non-zero temperature. Since the wavefunction of the novel quantum condensate that describes both superfluid and solid order is not established,  its physical manifestation is unclear. It is not unreasonable that DC superflow, which simple analogy with a conventional superfluid might lead us to expect, is not in fact supported by this ground state.  The supersolid is likely to possess properties not found in conventional superfluids. For example, the superfluid density of the hcp crystal may be anisotropic \footnote{This might lead to a sample geometry/orientation dependence of oscillator measurements \cite{Allen1982}}. And might a superfluid-like response be observable only at finite frequency? Decoherence of the quantum condensate by mobile dislocations in solid $^4$He could be important. In the light of the intriguing signatures already reported in rotation experiments \cite{Choi2010,Choi2012,Yagi2011}, the study of the interplay of angular oscillation at different frequencies with rotation may be a fruitful avenue of investigation. The theoretical grounds for the likely emergence of a coherent quantum condensate in a bosonic solid, recently reasserted \cite{Anderson2014}, continue to provide a strong motivation for this experimental search.

We would like to thank Dave Bosworth for making the torsional oscillator components and Alan Betts for
assembling them. We thank Jeevak Parpia for helpful discussions. Financial assistance from EPSRC and
SEPnet are gratefully acknowledged.

% below follows the extra references needed for the SM 
\nocite{Cornell1981}
\nocite{Agnolet!}
\nocite{Greywall1977}

\bibliography{bibfile}

\begin{thebibliography}{60}
\bibitem [{\citenamefont {Agnolet}()}]{Agnolet!}%
  \BibitemOpen
  \bibfield  {author} {\bibinfo {author} {\bibnamefont {Agnolet}},\ }\href@noop
  {} {\ }\bibinfo {note} {{[51] recommends using the shear modulus of BeCu for
  coin silver, namely $5.3\times10^{10}$\,Pa. However from measurements on
  oscillators such as the one used here, we have determined the value of the
  coin silver shear modulus to be $3.3\times10^{10}$\,Pa} }\BibitemShut
  {NoStop}%
\end{thebibliography}

%\bibliography{/Applications/bib}
\end{document}

\title{SUPPLEMENTARY INFORMATION:\\Frequency Dependence of the Supersolid Signature in Polycrystalline $^4$He}

\author{George Nichols}
\author{Malcolm Poole}
\author{Jan Ny\`eki}
\author{John Saunders}
\author{Brian Cowan}

\affiliation{Royal Holloway University of London, Egham, Surrey, TW20 0EX, UK.}%Lines break automatically or can be forced with \\

\maketitle

\section{Effects of stiffening of the Helium upon the Torsion Constants}
In this section we consider the effect of the complex shear modulus stiffening upon the effective torsion constants. 
   In the calculations that follow we have taken an 8\% stiffening,  corresponding to the values for which the complex measurements are available [18].

\subsection{Beamish effect}

In oscillator designs, such as ours, where the sample chamber is filled through a hole in the torsion
rod, the helium shear modulus will contribute to the rod's effective torsion constant. This can have
a significant effect, as pointed out by  Beamish \emph{et al}. [26] (Beamish effect).

We denote the radius of the torsion rod by $r_\mathrm o$ and the radius of the fill hole by $r_\mathrm
i$. If the shear modulus of the solid helium changes by the complex $\Delta \mu_\mathrm{He}$ then the
fractional change of the rod's complex torsion constant is
\begin{equation}
\frac{\Delta k}{k}=\frac{1}{(r_\mathrm o/r_\mathrm i)^4-1}\frac{\Delta \mu_\mathrm{He}}{\mu_\mathrm m}
\end{equation}
 where $\mu_\mathrm m$ is the shear modulus of the rod material.

 Our oscillator has two torsion rods. For rod 1:  
$r_\mathrm o =  0.80$\,mm and $r_\mathrm i = 0.30$\,mm. For rod 2: $r_\mathrm o =  0.90$\,mm and $r_\mathrm
i = 0.15$\,mm. This gives
 \begin{equation}\label{Delta-ks}
 \begin{split}
\frac{\Delta k_1}{k_1}&= 2.02\times 10^{-2}\frac{\Delta \mu_\mathrm{He}}{\mu_\mathrm m}\\  
\frac{\Delta k_2}{k_2}&= 7.72\times 10^{-4}\frac{\Delta \mu_\mathrm{He}}{\mu_\mathrm m}.
\end{split}
\end{equation}
Changes in the torsion constants of both rods will contribute to changes of the oscillator period and
dissipation:
\begin{equation}\label{Delta_omegas}
\begin{split}
-\left(\frac{\Delta P_\mathrm{a}}{P_\mathrm{a}}+\frac{1}{2}\Delta\frac{1}{Q_\mathrm a}\right)&=(\tfrac{1}{2}-\alpha)
\frac{\Delta k_1}{k_1}+
\qquad\,\,\,\alpha \frac{\Delta k_2}{k_2} \\
-\left(\frac{\Delta P_\mathrm{s}}{P_\mathrm{s}}+\frac{1}{2}\Delta\frac{1}{Q_\mathrm s}\right)&=\qquad\,\,\,\,\alpha
\frac{\Delta k_1}{k_1}+
(\tfrac{1}{2}-\alpha) \frac{\Delta k_2}{k_2}
\end{split}
\end{equation}
where
\begin{equation}
\alpha=\frac{\partial\ln\omega_\mathrm s}{  \partial\ln k_1} =
\frac{(\omega_\mathrm a^2-\omega_1^2)}{2(\omega_\mathrm a^2-\omega_\mathrm s^2) }
\end{equation}  
and the other coefficients of Eq~\eqref{Delta_omegas} follow from differentiating Eq.~(3).
Here we have denoted $\omega_1=\sqrt{k_1/I_1}$.

This gives contributions to the two mode period shifts and dissipation of
\begin{equation}
\begin{split}
-\left(\frac{\Delta P_\mathrm{a}}{P_\mathrm{a}}+\frac{1}{2}\Delta\frac{1}{Q_\mathrm a}\right)&= 1.51\times10^{-3}\frac{\Delta
\mu_\mathrm{He}}{\mu_\mathrm m}\\
-\left(\frac{\Delta P_\mathrm{s}}{P_\mathrm{s}}+\frac{1}{2}\Delta\frac{1}{Q_\mathrm s}\right)&= 8.97\times10^{-3}\frac{\Delta
\mu_\mathrm{He}}{\mu_\mathrm m}.
\end{split}
\end{equation}
We note that for both modes the dominant contribution is from rod~1.
%\begin{equation*}
%\left(
%\begin{split}
%-\frac{\Delta P_\mathrm{a}}{P_\mathrm{a}}&= (1.16+0.34)\times10^{-3}\frac{\Delta \mu_\mathrm{He}}{\mu_\mathrm
%m}\\
%-\frac{\Delta P_\mathrm{s}}{P_\mathrm{s}}&= (8.92+0.04)\times10^{-3}\frac{\Delta \mu_\mathrm{He}}{\mu_\mathrm
%m}.
%\end{split}
%\right)
%\end{equation*}

We take the  shear modulus of the coin silver rods to be $3.3\times10^{10}$\,Pa [55]. The shear modulus of solid $^4$He at 42 bar is approximately
 $1.5\times 10^7$\,Pa [16, 56].

%(On the assumption of a 100\% reduction of the helium's shear modulus, ${\Delta \mu_\mathrm{He}}/{\mu_\mathrm m}=4.5\times10^{-4}$ and then $\Delta P_\mathrm s= 10$\,ns and $\Delta P_\mathrm a= 0.34$\,ns). 

Taking the helium complex shear modulus measurements of Syshchenko \textit{et al.}~[18], with an 8\% stiffening,     then leads to the Cole-Cole plots 
shown as the blue and the
red curves marked B in Fig.~4; the horizontal placing is arbitrary. The real part, 
$\Re\Delta\mu_\mathrm{He}/\mu_\mathrm m=-3.6\times10^{-5}$,
gives to an increase in $\Delta P_\mathrm s$ of 0.8\,ns and an increase in $\Delta P_\mathrm a$ of
0.03\,ns.  
\subsection{Maris effect}

Maris [27] has suggested a further way in which a changing helium shear modulus can influence
the oscillator's response (Maris effect). The torsion of rod 2 transmits the torque to the helium specimen.
However the base of the sample chamber participates in this process and so the \emph{effective} torsion constant
should include this contribution. Maris pointed out that the helium in contact with the base of the chamber
will affect its stiffness. Thus a change of the helium shear modulus will cause a change in the base
plate contribution to the torsion constant $k_2$. 

Using a simple model, and supported by numerical simulations, Maris estimated the fractional change to
be 
\begin{equation}
\frac{\Delta k_2}{k_2}=\frac{r_\mathrm o^3}{8lw^2}\frac{\Delta\mu_\mathrm{He}}{\mu_\mathrm m}
\end{equation}
where $r_\mathrm o$ and $l$ are the diameter and length of rod 2,  and $w$ is the thickness of the cell
base. For our oscillator $l=2.50$\,mm and $w= 1.0$\,mm. 
Then for the geometry of our system this gives
\begin{equation}
\frac{\Delta k_2}{k_2}=3.65\times 10^{-2}\frac{\Delta\mu_\mathrm{He}}{\mu_\mathrm m}.
\end{equation}
Comparison with Eq.~\eqref{Delta-ks} shows that for this effect,  $k_2$ is some fifty times that caused
by the helium in the torsion rods.  

The influence on the period shift and dissipation is then, from Eq.~\eqref{Delta_omegas},
\begin{equation}
\begin{split}
-\left(\frac{\Delta P_\mathrm{a}}{P_\mathrm{a}}+\frac{1}{2}\Delta\frac{1}{Q_\mathrm a}\right)&= 1.61\times10^{-2}\frac{\Delta
\mu_\mathrm{He}}{\mu_\mathrm
m}\\
-\left(\frac{\Delta P_\mathrm{s}}{P_\mathrm{s}}+\frac{1}{2}\Delta\frac{1}{Q_\mathrm s}\right)&= 2.12\times10^{-3}\frac{\Delta
\mu_\mathrm{He}}{\mu_\mathrm
m}.
\end{split}
\end{equation}
The Cole-Cole plots for this are shown
as the blue and the
red curves marked M in Fig.~4; the horizontal placing is arbitrary. This effect results in  an increase in $\Delta P_\mathrm s$ of 0.20\,ns and
an increase in $\Delta P_\mathrm a$ of 0.30\,ns. These will have negligible effect on the inferred MMI. 

\subsection{Corrections to the MMI}

The result of these corrections is shown as the triangles in Fig.~3: $\Re\Delta I_\mathrm{eff}(\omega_\mathrm
a)/I_\mathrm{He}=(6.84\pm 0.37)\times10^{-3}$,
$\Re\Delta I_\mathrm{eff}(\omega_\mathrm s)/I_\mathrm{He}=(1.87\pm 0.16)\times10^{-3}$ and the intercept
of $(1.67\pm0.17)\times 10^{-3}$ is the remaining MMI fraction.
It is perhaps surprising that while  the Maris effect on $k_2$ is some fifty times greater than the Beamish
effect, its influence on the MMI is so much smaller. Essentially this is because, in
our compound oscillator, the MMI is predominantly sensitive to the symmetric mode period,
and this is relatively insensitive to changes in $k_2$.

\section{Complex Shear Modulus}
The complex shear modulus stiffening was observed by Syshchenko \textit{et al.} [18] at a pressure of 33.1~bar. Measurements were made at a range of frequencies between 0.5~Hz and 8~kHz. These measurements have been discussed by Su \textit{et al.} [50] in the context of a glass anomaly. 

\begin{figure}[h]
\includegraphics[scale=0.5]{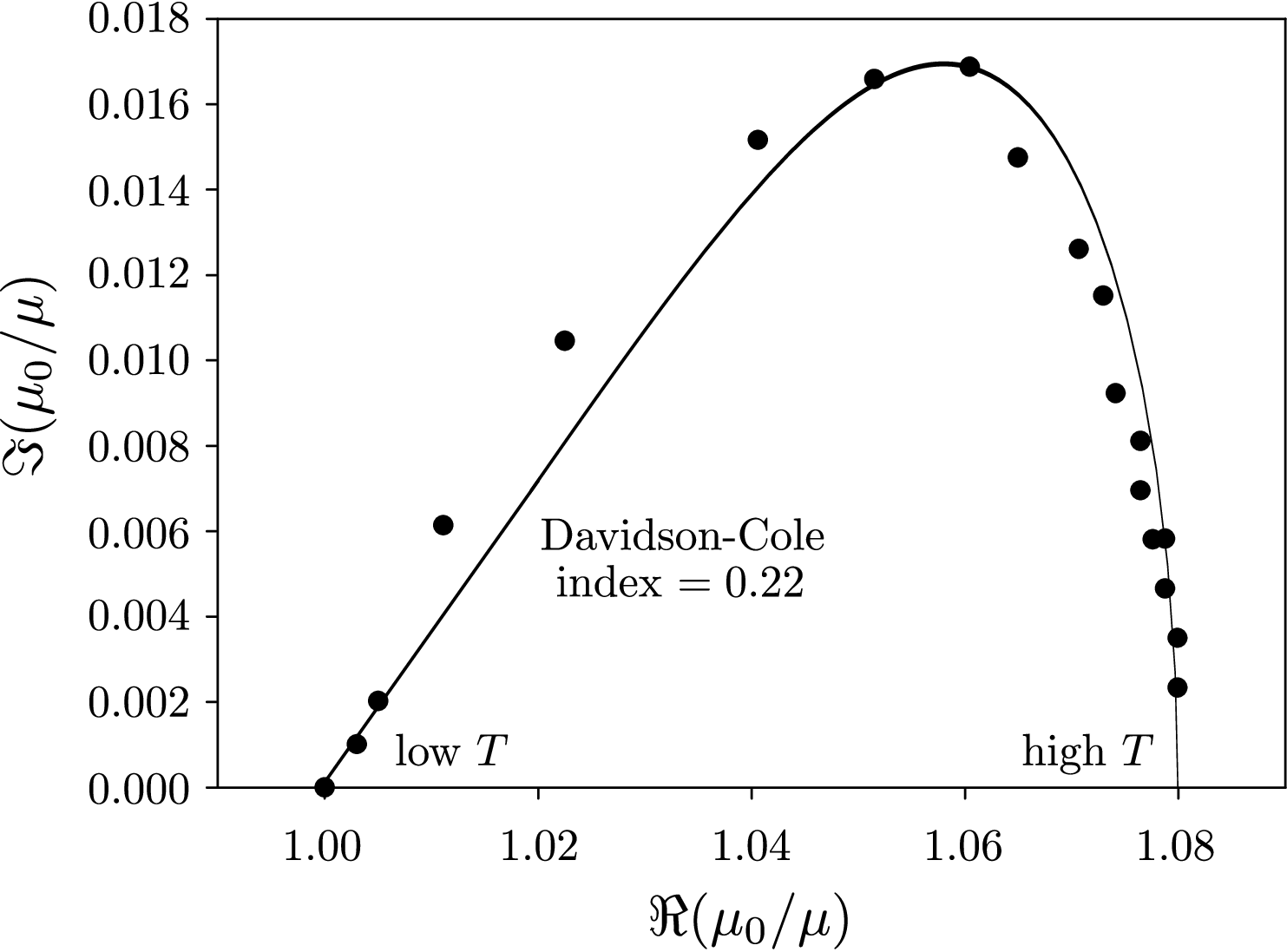}
\caption{\label{Syshchenko_DCplot} Complex plot of the inverse shear modulus data [18] showing the stiffening upon cooling.  }
\end{figure}

In Fig.~\ref{Syshchenko_DCplot} we have plotted some smoothed points from Syshchenko's 200~kHz measurements. And we have shown a Davidson-Cole function, with an index $\gamma=0.22$, corresponding to the observed aspect ratio of the data. This is the functional form of the complex shear modulus we have used in the above analysis.   

We note that  Su \textit{et al.} give a complex plot of all of the Syshchenko data. The density of the points together with the noise scatter results in a figure whose asymmetry cannot be clearly discerned and for this reason they use a Cole-Cole fitting function. However this distinction is immaterial for the above analysis.   

\section{Drive Dependence}
\begin{figure}[h]
\includegraphics[scale=0.5]{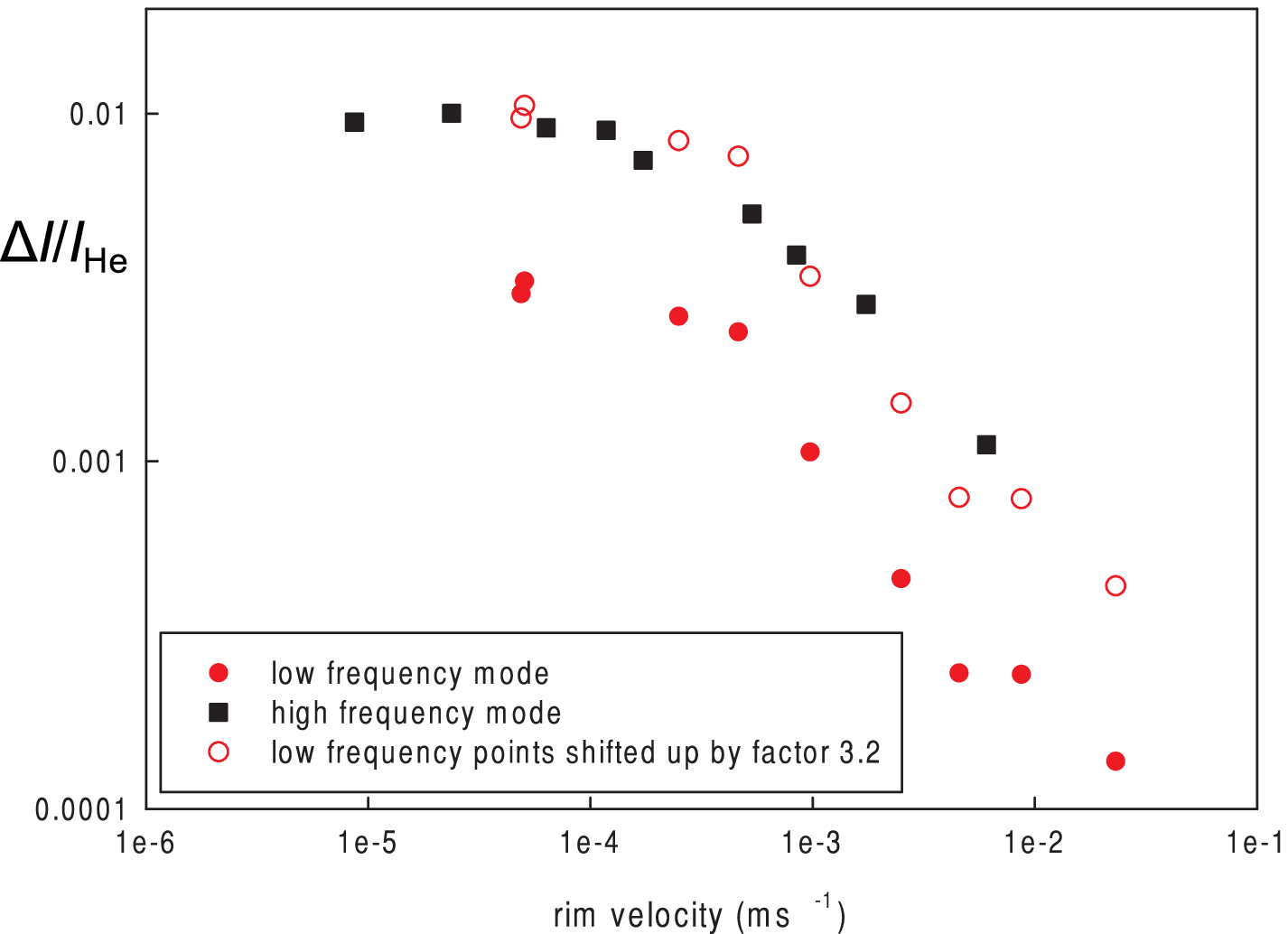}
\caption{\label{drivedep} Velocity dependence of MMI fraction.}
\end{figure}

The oscillator measurements at different drive levels shown in Fig.~2  permit 
 inference of the drive dependence of the missing moment of inertia. From each curve of the figures the corresponding MMI fraction was determined, as described in the text. These are plotted against rim velocity in Fig.~\ref{drivedep}. The rim velocities were determined as follows.

The signal level at the preamplifier ouput $V_\mathrm{out}$ is proportional to the velocity of the oscillator's vanes $v_\mathrm{vanes}$:
\begin{equation}
V_\mathrm{out}=\frac{G_\mathrm AV_\mathrm{DC}C^2}{\varepsilon_0A} v_\mathrm{vanes}
\end{equation}
where $G_\mathrm A$ is the gain (transresistance) of the preamplifier, $V_\mathrm{DC}$ is the voltage of the DC bias supply, $C$ is the electrode capacitance and $A$ its area, $\varepsilon_0$ is the permittivity of free space. However the head and the vanes of the two-mode oscillator move with different angular speeds. In particular the head moves faster than the vanes in the low frequency mode and slower (and anti-phase) in the high frequency mode. If the specimen's rim is at radius $r_\mathrm{rim}$ and the radius of the vane is $r_\mathrm{vane}$ then the velocity at the rim, $v_\mathrm{rim}$ is given by       
\begin{equation}
v_\mathrm{rim}=v_\mathrm{vane}\frac{1}{1-\omega_\mathrm{a/s}^2/\omega_2^2}\times\frac{r_\mathrm{rim}}{r_\mathrm{vane}}.
\end{equation}

The MMI fractions are plotted against the thus-determined rim velocities in Fig.~\ref{drivedep} for the two mode frequencies. Although there is some `noise' in the low frequency points, the data from the two modes behave similarly; they roll off with a similar velocity dependence. This is emphasised by shifting the low mode data up and over the high frequency points. The data are consistent with a critical velocity of the order of  $100\,\mu\mathrm{ m}\,\mathrm{s}^{-1}$ although it should be borne in mind that with our pancake geometry the inner regions will be moving slower than the rim.

The similar behaviour of the high and low frequency points indicate that the phenomenon is a velocity effect rather than an acceleration or  displacement effect, in agreement with the contention of Aoki and Kojima [14]. However it contrasts with the critical amplitude (strain) effects observed in studies of the nonlinear elastic response [17].